# Selective crystal growth of indium selenide compounds from saturated solutions grown in a selenium vapor

Chao Tang*[1,2], Yohei Sato[1], Katsuya Watanabe[1], Tadao Tanabe[1,3], Yutaka Oyama[1]

[1] *Department of Materials Science, Graduate School of Engineering, Tohoku University*

[2] *Research institute of electrical communication, Tohoku University*

[3] *Graduate School of Science and Engineering, Shibaura Institute of Technology*

**Abstract**: Indium selenide compounds are promising materials for energy conversion, spintronic applications, and chemical sensing. However, it is difficult to grow stoichiometric indium selenide crystals due to the high equilibrium selenium vapor pressure and the complicated phase equilibrium system of indium selenide compounds. In this paper, we apply a novel and convenient crystal growth method, in which the saturated solution is grown by the application of a selenium vapor. In this method, selenium vapor at a pressure higher than the saturated vapor pressure is applied to molten indium, such that the selenium continuously dissolves into the indium solution until the solubility at the growth temperature is reached; then a slow cooling process results in crystallization. The selenium dose is matched to its solubility so that indium selenide compounds can be selectively grown just by controlling the temperature without the need to consider the chemical ratio in the solution. The pressure of the vapor is controlled during the whole growth process so that any deviation from a stoichiometric composition in the crystal can be controlled. Both InSe and $In_2Se_3$ were successfully grown using this method, and we have investigated the growth mechanism and found the growth window for each compound. This study presents a novel and convenient approach to fabricating transitional metal chalcogenides without deficiencies in the chalcogen element.

* Corresponding Author: tang.chao.c4@tohoku.ac.jp

Research Institute of Electrical Communication, Tohoku University. 2-1-1 Katahira, Aobaku, Sendai, 980-8577 Japan

## 1. Introduction

Transition-metal monochalcogenide (TMM) materials have been attracting increasing attention owing to their favorable properties [1-4]. Indium selenide compounds are typical TMMs, and have a narrow bandgap compared with transition metal sulfides and gallium compounds. The indium selenide compounds attracting most attention are InSe, $In_2Se_3$, and $In_4Se_3$. The bandgap energy of InSe and $In_2Se_3$ is about 1.3 eV, which makes them potential candidates for the next-generation of solar energy harvesting devices [5, 6] and highly-sensitive infrared detectors [7-9]. Also, InSe is considered to be a possible candidate for exploring the frontiers of spintronics [10, 11]. Moreover, InSe can be used for the adsorption of carbon oxide [12, 13]. Furthermore, $In_4Se_3$ has been reported as being a high-performance thermoelectric material with a considerable Seebeck coefficient of 1.48 [13].

Up to now, indium selenide crystals have been usually prepared by the Bridgman technique [6, 14-26]. The Bridgman technique has the following disadvantages, which limit the crystallinity of the crystals. Firstly, crystals are grown from a molten mixture of In and Se where the temperature is much higher than the melting point of the growing crystal. Thermal motion of the molecules is induced by defects in the crystal [27]. Also, the vapor pressure of selenium during growth is high so that saturated selenium evaporates from the melt during growth, causing the composition of the crystal to be non-stoichiometric [28, 29]. Hence, the ratio of the elements in the melts used in the Bridgman method always differ from the stoichiometry of the target compound itself [30]. As a result, Bridgman-grown crystals are commonly selenium deficient. To overcome these problems, our group attempted to grow TMMs using a selenium vapor at a controlled pressure at low temperature [31, 32]. However, the driving force for growth was only supplied by the supersaturation induced by the concentration gradient and diffusion. Consequently, the growth speed was limited [33]. The crystal growth of transition metal phosphides using this technique have also been reported [34]; however, this research was limited to III-V compounds.

Here, we introduce an improved method based on gradual cooling, in which the TMM crystals are grown from a supersaturated solution formed by controlling the pressure of the chalcogen vapor. This method has the following advantages compared with the existing technique. Firstly, the pressure of the chalcogen vapor is controlled during growth thus maintaining the stoichiometry. Secondly, the growth rate is relatively high compared with that in the previous study [31]. Finally, the chemical composition of the solution depends only on temperature, so, by controlling the growth temperature, different specific compounds in the phase diagram can be precisely targeted [21]. In this study, we fabricated high-quality InSe and $In_2Se_3$ to demonstrate this method. We found the growth window for each compound and investigated the underlying mechanism.

## 2. Crystal Growth

We conducted crystal growth experiments using the different growth conditions listed in Table. 1. Let us take the successful growth of InSe and In$_2$Se$_3$ (conditions I and II) as examples to explain the growth procedure. A schematic diagram of the growth equipment is shown in Fig. 1. Ten grams of high purity selenium (5N, Kojundo Chemical Lab. Co., Ltd.) is placed at the bottom of the quartz ampoule. Then a 0.5 mm diameter quartz rod, which is smaller than the inside diameter of the quartz ampoule, is placed above the selenium ingots. The crucible, with 4 g of indium (6N, Dowa Co., Ltd.), is on top of the quartz rod. The ampoule is sealed after evaporation down to $10^{-6}$ Torr. Then the ampoule is placed in a vertical electrical furnace that has four separate temperature-controlled regions. Fig. 1a and 1b show the temperature profiles in the furnace during the growth of the crystals. The temperature at the crucible ($T_G$) and the bottom of the ampoule ($T_V$) are adjusted to control the growth temperature and the selenium vapor pressure. The equilibrium vapor pressure of pure selenium ($P_V$) is given by Baker's equation [35],

$$\log P_V \, (atm) = -\frac{5043}{T_V(K)} + 5.265 \tag{1}$$

The gap between the quartz rod and the ampoule is sufficiently narrow for the pressure of the selenium vapor at the crucible ($P_G$) to be described by the equation given by Nishizawa [36].

$$P_G = P_V \sqrt{\frac{T_G}{T_V}} \tag{2}$$

The crystal growth procedure is shown in Fig. 2a. There are four steps in the growth process. The growth temperature ($T_G$) with respect to the growth time is shown in Fig. 2b, and the temperature ($T_V$) and pressure of the selenium vapor ($P_G$) with respect to the growth time are shown in Fig. 2c. Firstly, the temperature of the crucible is rapidly raised to a temperature higher than the growth temperature of InSe (Step I). Then the temperature is maintained for 10 hours to homogenize the solution (step II). Then we increase $T_V$ to apply the specified pressure of the selenium vapor and this is maintained for 12 hours; the selenium vapor dissolves into the In solution until the solution becomes saturated at $T_G$ if the pressure of the selenium vapor is higher than the equilibrium vapor pressure of the saturated solution at $T_G$ (Step III). Finally, the crucible is gradually cooled at a rate of 1 °C/hour as shown in the inset of Fig. 2b, and the InSe crystal grows from the supercooled solution (Step IV). The pressure of the selenium vapor is maintained during step IV to prevent selenium evaporating from the solution. The whole system is cooled to room temperature after growth for 24 h. Likewise, the temperature and pressure profiles for the $In_2Se_3$ growth are shown in Fig. 2e and f.

## 3. Result and Discussion

Crystal growth was conducted at many different pressures and temperatures, and these are listed in Table. 1. The most promising conditions for the growth of InSe and $In_2Se_3$ are conditions I and II. In Fig. 3 we show the morphologies of an as-grown crystal of InSe. Fig. 3a shows the bottom of the ampoule, from where the pressure of the vapor is controlled, after growth. Some selenium remains, which means not all of this was used up during growth. A metallic looking bulk material with a length of about 3 cm was obtained from the crucible (Fig. 3b). The surface of this has stripes in one direction (Fig. 3c), indicating the direction of the (00n) plane [37]. Due to the random nucleation, each sample grows in an arbitrary direction compared to our previous samples grown by the temperature difference method [31]. There are some bulges at the top of the ingot, which quickly react with hydrochloric acid, and these bulges were confirmed by inductively coupled plasma mass spectrometry (ICP-MS) to be pure selenium. This Se solidified when the crucible was cooled down to room temperature. After dissolving the metal in hydrochloric acid, a mirror-like surface was exposed (Fig. 3e). Microscope images of the surface are shown in Fig. 3e-h. The layer-stacking morphology can be observed in Fig. 3e. The six-fold symmetry of the surface structures are shown in Figs. 3f and g, and a terrace formed during growth is apparent in Fig.3h.

The crystal phases of the as-grown crystals were confirmed by x-ray diffraction (XRD) using a D8 ADVANCE (Bruker AXS Co., Ltd.). XRD spectra taken from different parts of the bulk crystal are shown in Fig. 4. For the XRD pattern of the sample grown using condition I, only the peaks with Miller indices of (00n) (n = 3, 6, 9 ….) can be observed, which indicates the sample is InSe with R3m symmetry [38]. The lattice constants, a = b = 4.1 Å and c = 24.9 Å, were derived from the strongest (006) peak at $2\theta = 21.3°$. The XRD spectra of the sample taken 3, 6, and 9 mm from the bottom of the ingot are shown in Fig. 4b. Similarly, only the (00n) peaks can be seen in these spectra. The peak intensity decreases moving away from the bottom of ingot, for which we give the following explanation. The pressure of the selenium vapor is greater than the equilibrium pressure and this is continuously applied to the liquid surface during growth. The saturated selenium diffuses from the surface of the liquid towards the bottom of the crucible. The crystals are grown during cooling and the supersaturation increases as the temperature decreases. Consequently, there is excessive selenium in the solution. Unable to escape from the surface of the liquid, the selenium accumulates in the upper part of crucible. so that the crystal becomes indium deficient towards the top of the crystal, thereby affecting the peak intensity. The relationship between the full width at half maximum (FWHM) of the (0 0 15) peak and the distance from the bottom of the sample where the peaks were observed is shown in Fig. 4c. The FWHM increases with the distance, which confirms that the crystallinity has been affected by the indium deficiency. The XRD pattern of the sample grown under condition II is shown in Fig. 4d, indicating the sample is $In_2Se_3$ with R3m (No. 166) symmetry [39]. Only (00n) peaks were observed in the XRD pattern, which indicates the sample is single crystal. The lattice constants, a = b = 4.1 Å and c = 28.8 Å, were derived from the strongest (006) peak at 2θ = 18.3°. XRD mapping from the bottom to 9 mm above the bottom of the bulk crystal is shown in Fig. 4e. These results suggest that the material is single crystal $In_2Se_3$ from the bottom up to 1 cm from the bottom. As shown in Fig. 4f, the FWHM of the (0 0 15) peak taken from the bottom of the bulk crystal is smaller than those taken from higher up. It is considered that the inhomogeneity of the selenium in the solution affects the growth of both InSe and $In_2Se_3$.

The backscattered Raman spectra were taken using an NRS-5100 (JASCO Corp.) to confirm the polytypes of the as-grown samples. With excitation at 532 nm, the Raman spectra of samples grown under conditions I and II are shown in Fig. 5a and b, respectively. In the Raman spectrum of the sample grown under condition I, the transverse optical phonon $A_1$(TO) mode at 225 cm$^{-1}$ verified that the sample is γ-InSe. The Raman modes E (176 cm$^{-1}$) and $A_1$(LO) (115 cm$^{-1}$) can also be seen in the spectrum. In the Raman spectrum of the sample grown under condition II, the E mode at 187 cm$^{-1}$ and the strong $A_1$(TO) mode at 182 cm$^{-1}$ indicate the material is α-$In_2Se_3$ [40]. Moreover, the longitudinal optical phonon $A_1$(LO) mode at 244 cm$^{-1}$ [41] and the $A_1$(LO+TO) mode at 104 cm$^{-1}$ can also be seen in the spectrum. Raman spectra were also taken from different parts of the bulk crystal, showing modes only belonging to α-$In_2Se_3$; so the crystal grown under condition II was confirmed as single crystal α-$In_2Se_3$.

Because of the high vapor pressure of selenium, it is considered that Bridgman-grown indium selenide compounds are always non-stoichiometric. Here we measured the stoichiometry of the as-grown samples using inductive coupled plasma mass spectrometry (ICP-MS, Agilent 8800, Agilent Technologies, Ltd.). The results are shown in Table 1. With growth conditions I and II, the deviation from stoichiometric InSe and In$_2$Se$_3$ is less than 0.2 atomic percent. This is far smaller than a typical sample grown by the Bridgman technique [8, 37]. With the successful growth of InSe (conditions I, III, VI) and In$_2$Se$_3$ (conditions II, VII, VIII), it is expected that selenium vapor at a higher applied pressure will make the chemical ratio approach the stoichiometric ratio.

To explain these results, we propose a possible model in which the concentration of the supersaturated solution and the equilibrium vapor pressure are considered. The relationship between the concentration and the saturated vapor pressure can be described by Henry's law [42]:

$$P_{Se} = k c_{Se} \qquad (3)$$

Here, $P_{Se}$ and $c_{Se}$ are the saturated vapor pressure and the concentration of Se, respectively. $P_{Se}$ at a concentration of 60 at% is given by previously published results [43],

$$log\, P_{Se}\ (Torr, c_{Se} = 60\ at\%) = (8.056 \pm 0.96) - \frac{4949 \pm 170}{T_S(K)} \qquad (4)$$

where $T_S$ is the temperature of the saturated solution. The saturated selenium vapor pressure under different temperatures and concentrations can be obtained from Eqs. 3 and 4. $P_{Se}$ as a function of $c_{Se}$ and $T_S$ is shown in Fig. 6a. If the pressure of the selenium vapor is higher than $P_{Se}$, the selenium will dissolve into the indium solution until the solution becomes saturated, or until the solution reaches the concentration where the pressure of the selenium vapor is equal to $P_{Se}$. Because the system has already reached equilibrium, the selenium will not continuously dissolve into the solution even if it is not saturated. In the first situation, the crystal will be grown when slow cooling is started. Otherwise, crystals will not grow because the selenium concentration in the solution has not reached the saturated concentration. The temperature dependence of $P_{Se}$ when $c_{Se}$ is equal to 60 at% and 50 at%, indicating the stoichiometric ratios of InSe and $In_2Se_3$ are shown in Fig. 6b. The crystal growth conditions from I to IX are marked in the figure. $P_G > P_{Se}$ is satisfied with conditions I, II, and III, so samples with good crystallinity were obtained using these growth conditions. In contrast, with growth conditions IV, V, and IX, crystallization did not occur.

Note that the solution gradually cools down from the growth temperature over a period of 40 hours. The pressure of the selenium vapor may become larger than the vapor pressure at a specific temperature during cooling, so even when the pressure of the selenium vapor was smaller than $P_{Se}$, as with growth conditions VI and VIII, crystals were successfully grown. However, because there was insufficient time for the selenium to dissolve into the solution, the as-grown crystals were selenium deficient. On the other hand, $P_{Se}$ is a linear approximation, so with growth condition III, it is considered the chemical ratio of the as-grown crystal differs slightly from stoichiometric InSe.

We have divided Fig. 6b into four areas, separated by the temperature of the liquid phase line between InSe and $In_2Se_3$ in the phase diagram [21]. The crystallization zone of InSe and $In_2Se_3$ is in the area of high $P_{Se}$. The growth windows of InSe and $In_2Se_3$ are indicated in the figure. Below the crystallization zone is the transitional zone, where the selenium can dissolve into the solution during the gradual cooling, and stoichiometric crystals can be grown. In the non-crystallization region, crystals cannot be grown due to the low pressure of the selenium vapor. Because the gradual cooling is conducted in a temperature zone within the liquid phase line in the In-Se phase diagram, only the single crystal of that component will grow in the crystallization zone. By controlling the growth temperature ($T_G$) and the selenium vapor temperature ($T_V$) only, one can selectively grow indium selenide compounds without considering the selenium dose.

## 4. Conclusions

In this study, we introduced a novel crystal growth method for chalcogenide compounds, in which a saturated solution is produced by the application of the vapor of a chalcogen element, the pressure of which is controlled during the whole growth procedure, such that the growth temperature of this method is much lower than that of melt growth. Thus, the growth of non-stoichiometric crystals can be suppressed. Also, one can prepare the target crystal just by controlling the temperature without the need to consider the chemical composition of the solution in the crucible. We have successfully grown InSe and $In_2Se_3$ using this method. Moreover, we proposed a model to explain the growth mechanism and provide the growth windows of InSe and $In_2Se_3$. This method has the potential to be extended to the crystal growth of other transition metal chalcogenide compounds and benefit the preparation of samples for research into layered transitional metal chalcogenides.

**Acknowledgments**

This study is supported by Japanese society for promotion of science (JSPS) KAKENHI Grant number JP18J11396, JP19J20564, 20H02406.

**FIGURES**

**Figure 1.** (a) and (b): The temperature distributions for crystal growth conditions I and II (for the growth of InSe and $In_2Se_3$ respectively). (c): Schematic diagram of the quartz ampoule used in growth. The growth temperature ($T_G$) and the temperature of the selenium vapor are controlled according to the steps in the growth procedures.

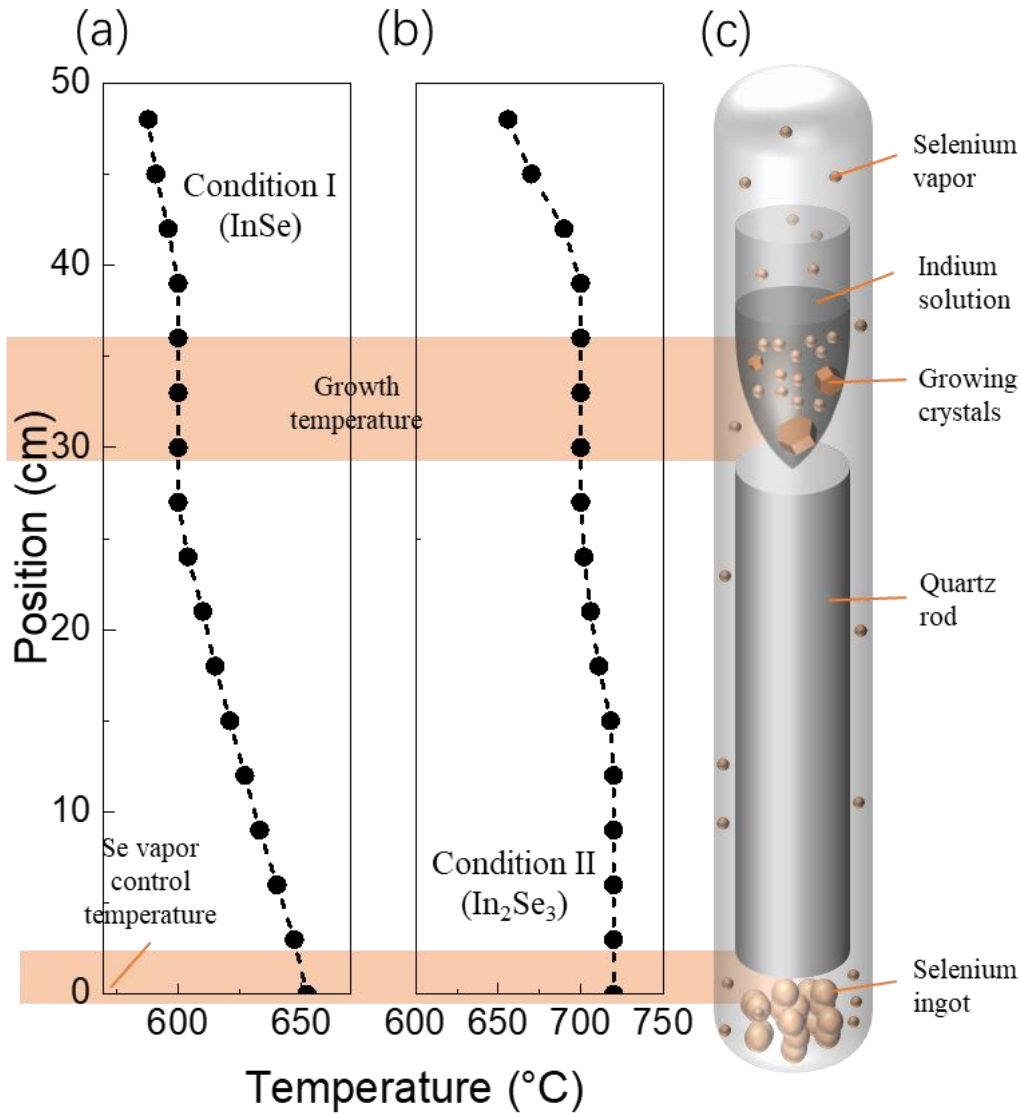

**Figure 2**. (a): Schematics of the four steps in the crystal growth procedure, I: setting the indium ingot in the crucible, II: melting and homogenizing the indium solution. III: application of the selenium vapor until the solution becomes saturated. IV: gradual cooling and crystal growth. (b): The growth temperature as a function of growth time for condition I. (Insert: The temperature during gradual cooling, the slope shows the cooling rate) (c) The time-dependent Se vapor control temperature and the pressure of the Se vapor in condition I. (d) and (e): the time-dependent $T_G$, $T_V$ and $P_G$ for growth condition II.

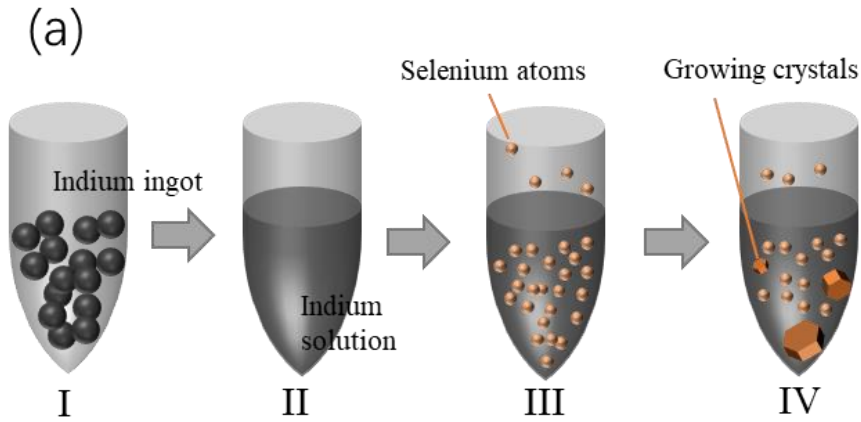

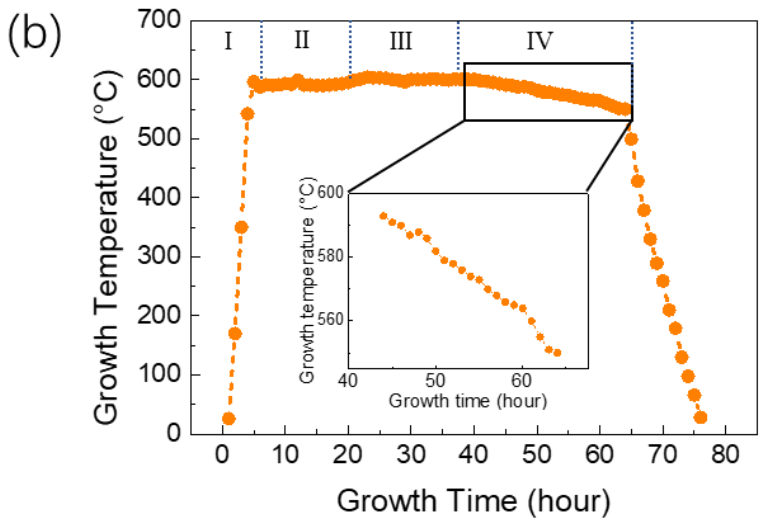

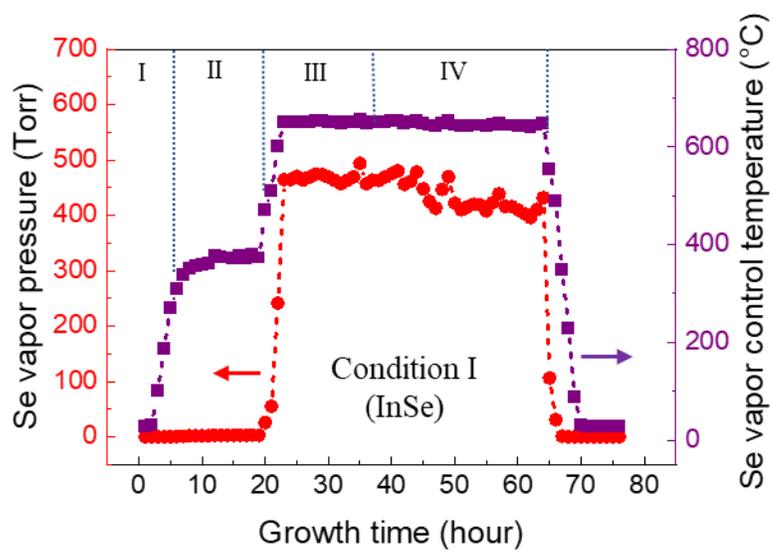

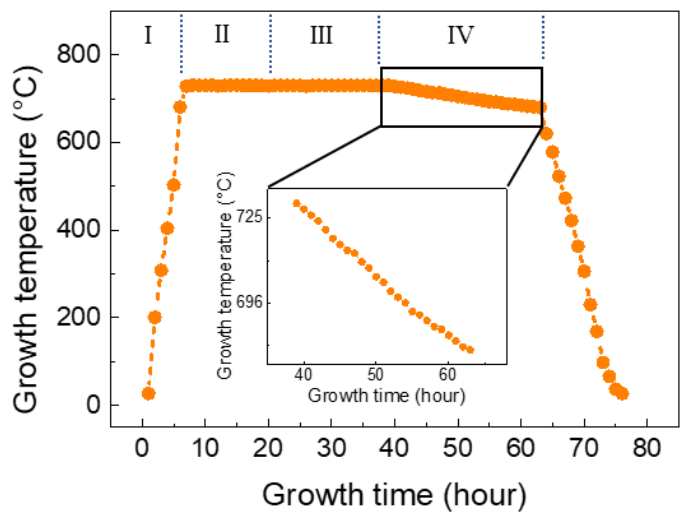

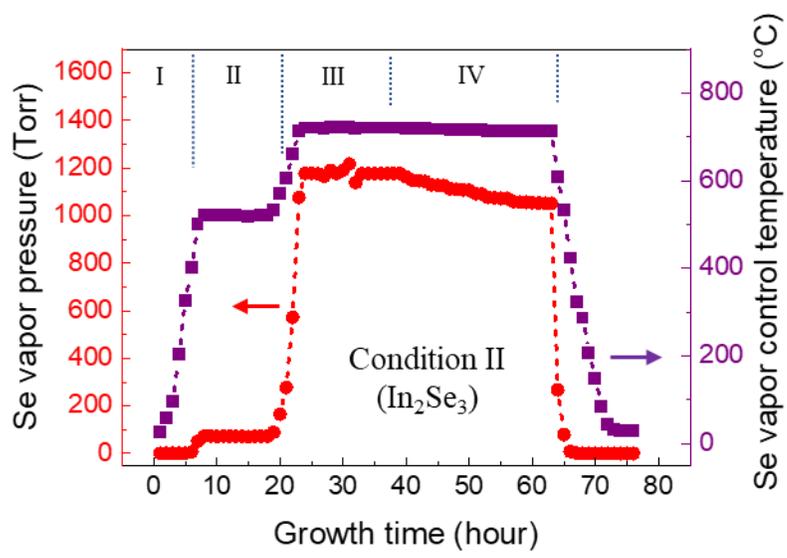

**Figure 3**. The morphology of the as-grown crystal grown using growth condition I, (a): The part of the ampoule from where the pressure of the selenium vapor is controlled. Some Se remains after growth. (b): The crucible after growth. (c): The ingot taken from the crucible. Condensed Se formed during cooling to room temperature remains at the top, below is the crystal. (d) Microscope image of the topside of the bulk crystal after removing the Se with hydrochloric acid. (e), (f) and (g): Microscope image of the surface perpendicular to the c-axis. Growth terraces can be seen in (g).

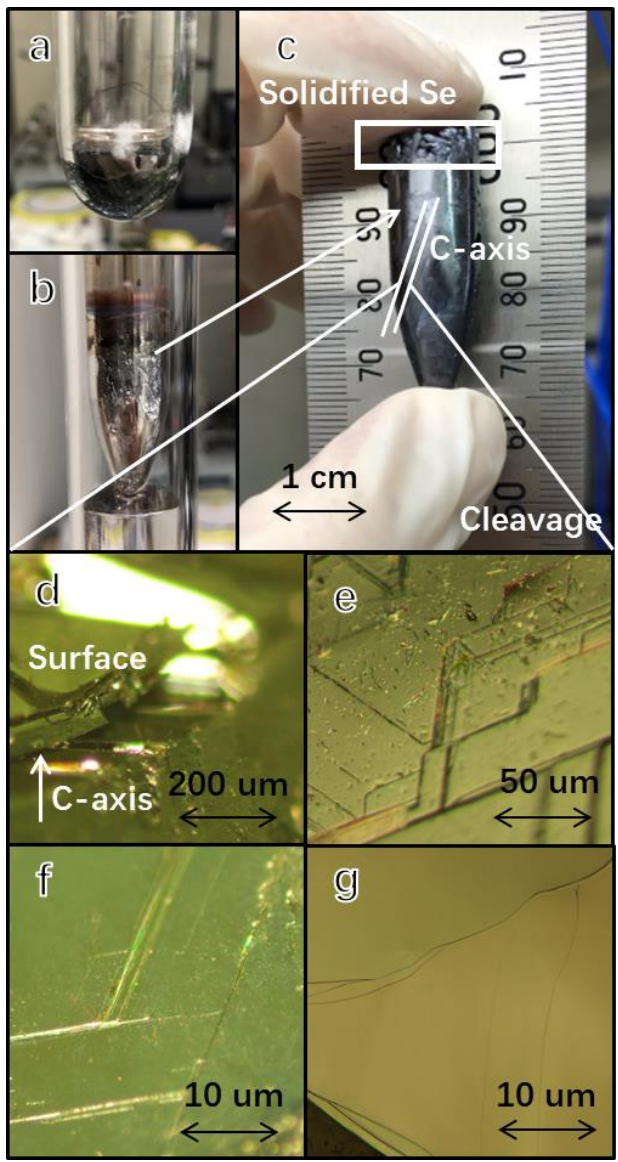

**Figure 4**. XRD patterns of as-grown crystals. (a): at the bottom of the bulk crystal grown using condition I (InSe). (b): at 3 mm, 6 mm, 9 mm from the bottom of the same sample. (c): the intensity and full width at half maximum (FWHM) of the (00 15) peak as a function of the distance from the bottom of the ingot. (d), (e) and (f): corresponding results for the crystal grown using condition II ($In_2Se_3$).

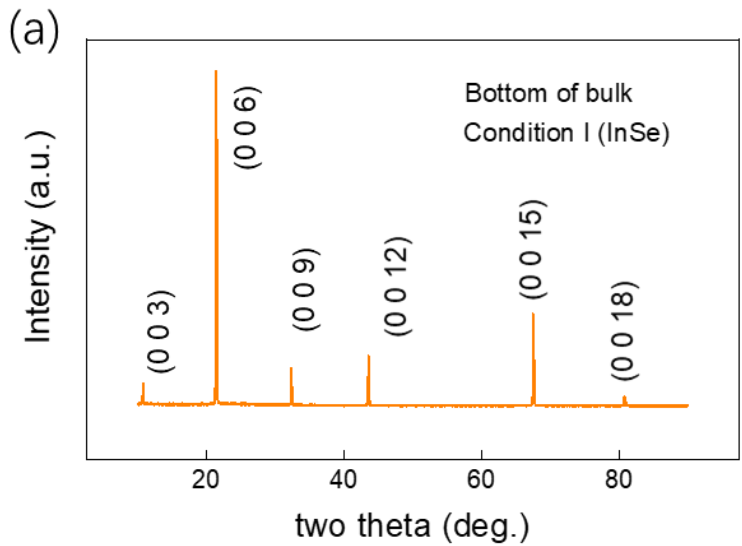

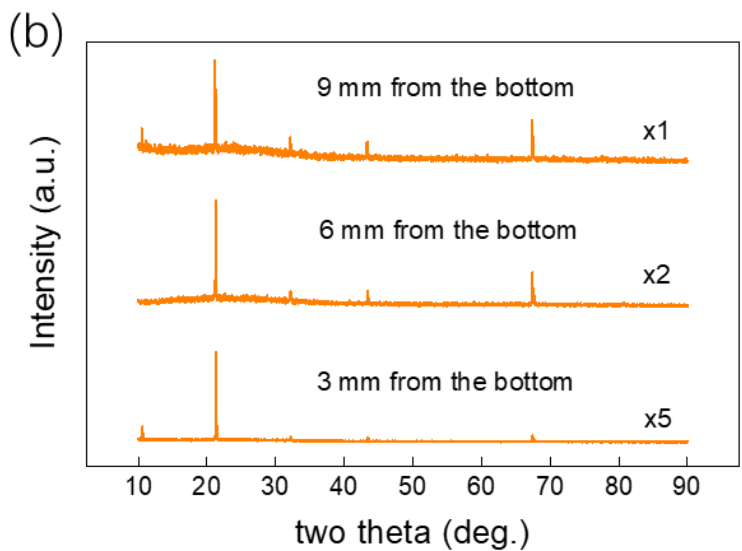

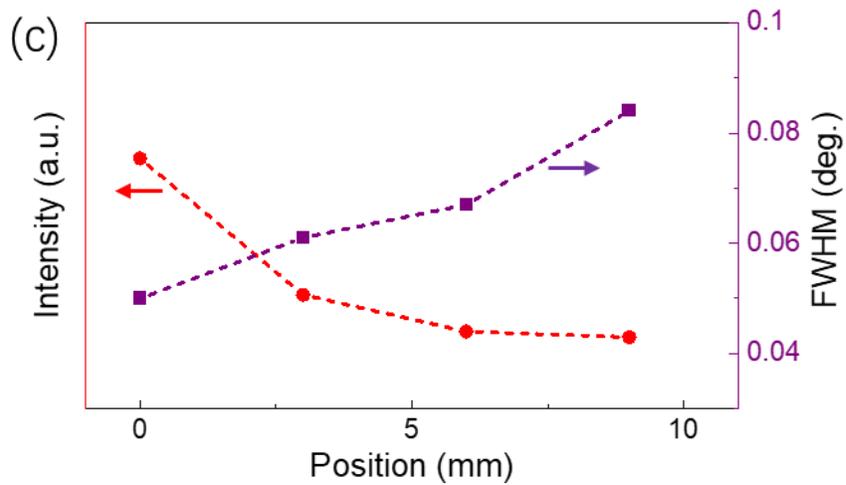

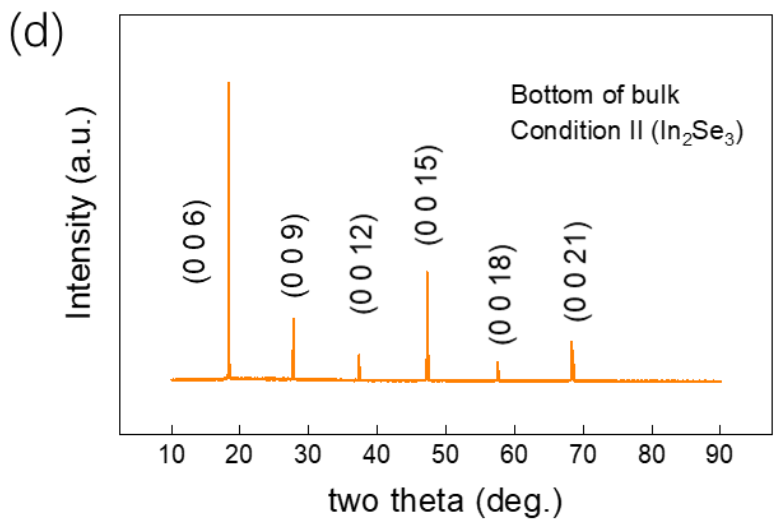

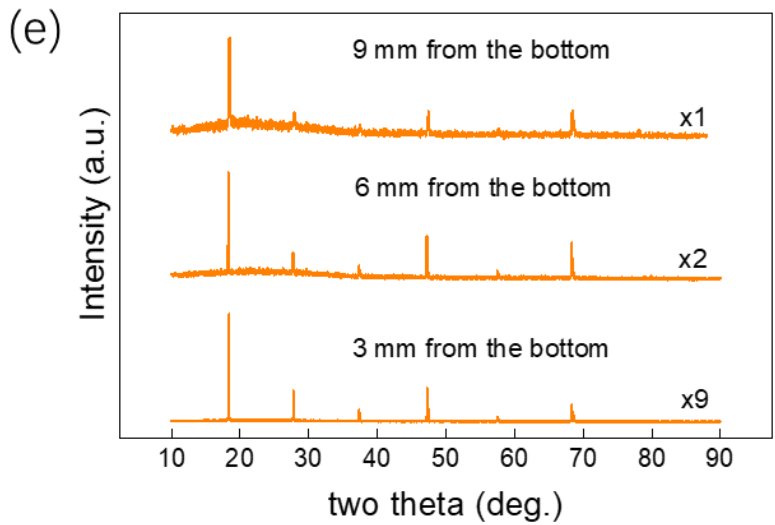

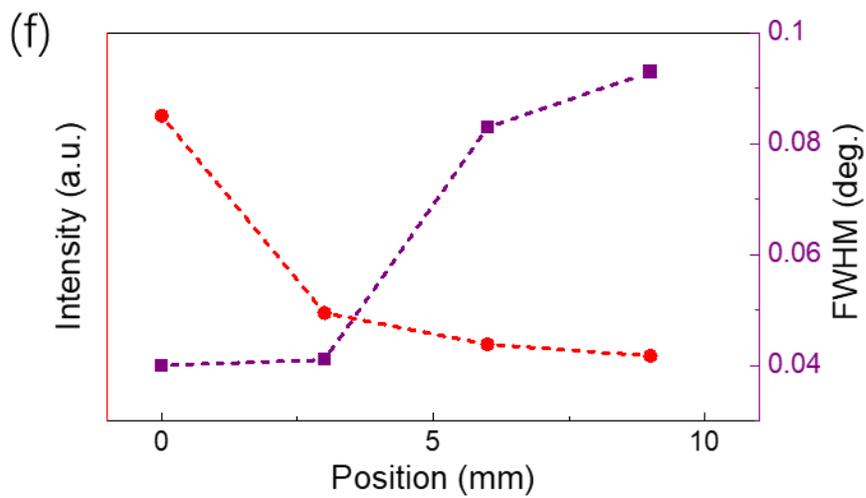

**Figure 5**. Backscattered Raman spectra of the sample grown using conditions I (a) and II (b)

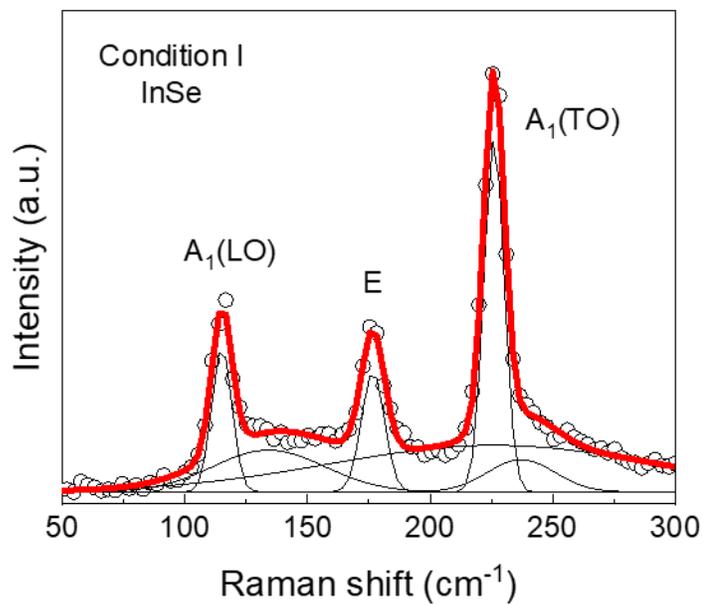

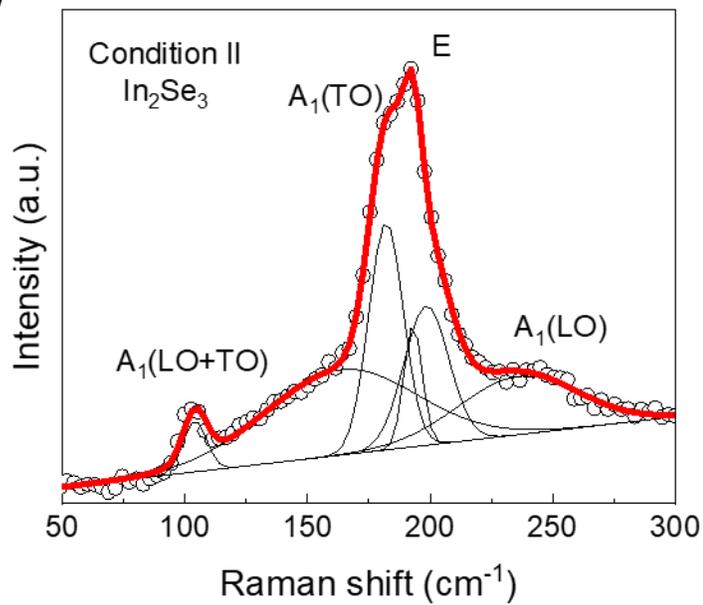

**Figure 6.** (a) the equilibrium selenium vapor pressure for different concentrations ($P_{Se}$) as a function of temperature. (b) The growth window for indium selenium compounds at different Se vapor pressures and temperature: the dashed line is the $P_{Se}$ for 50 at% and 60 at% selenium concentrations. The growth experiments are marked by blue dots. The yellow part shows the transitional zone and the red region is the non-crystallizable zone.

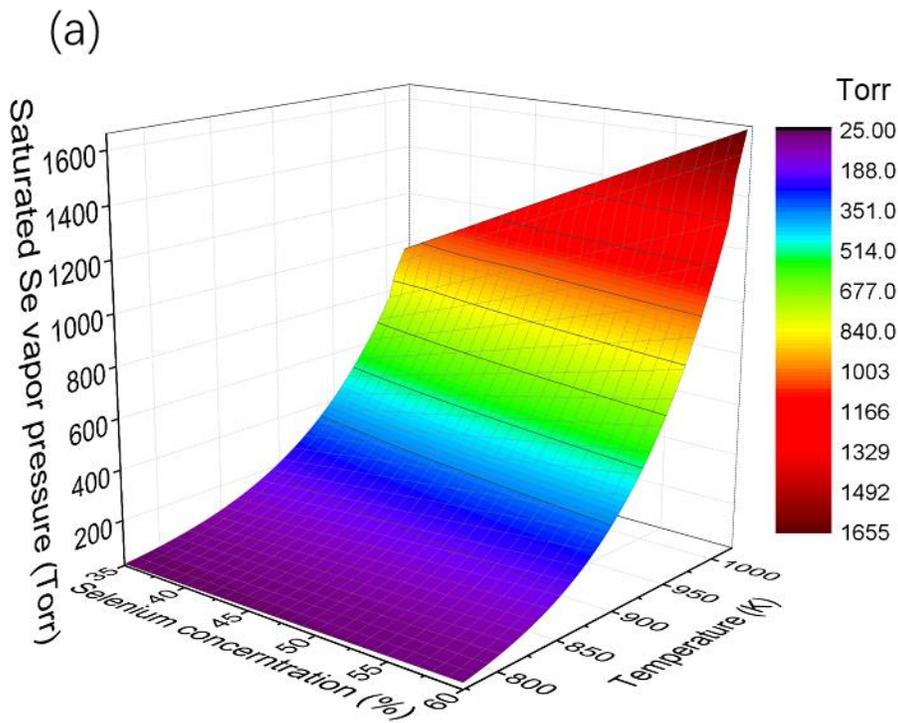

(b)

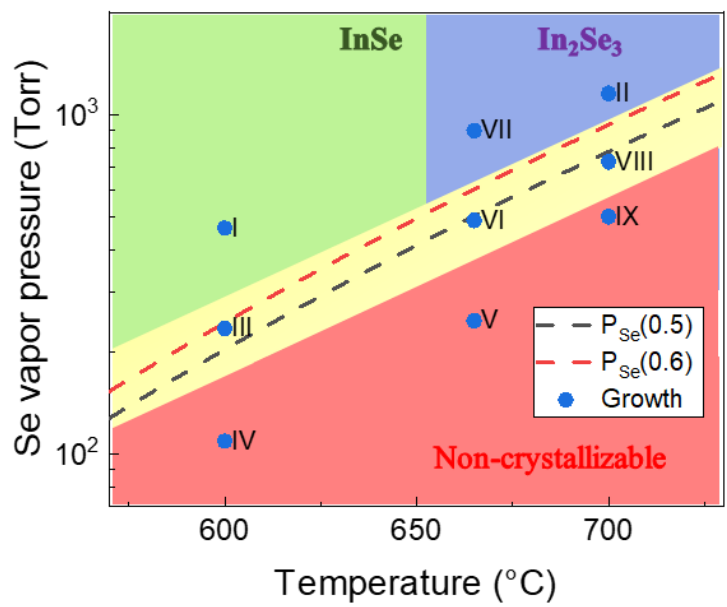

**TABLES.**

**Table. 1.** The growth conditions, as-grown crystal phases, and the stoichiometric properties of indium selenide compounds. The phases of the as-grown crystals were confirmed by Raman spectroscopy and XRD; the chemical composition of the indium and selenium were measured by ICP-MS.

| No. | $T_G$ (°C) | $T_V$ (°C) | $P_V$ (Torr) | $P_G$ (Torr) | Crystal | In (at%) | Se (at%) |
|---|---|---|---|---|---|---|---|
| I | 600 | 650 | 482 | 463 | **InSe** | 50.0 | 50.0 |
| II | 700 | 720 | 1170 | 1153 | **In$_2$Se$_3$** | 40.1 | 59.9 |
| III | 600 | 600 | 235 | 235 | **InSe** | 50.5 | 49.5 |
| IV | 600 | 550 | 105 | 109 | **In** | 100.0 | 0.0 |
| V | 665 | 600 | 235 | 247 | **In** | 82.3 | 17.7 |
| VI | 665 | 650 | 482 | 487 | **InSe** | 50.3 | 49.7 |
| VII | 665 | 700 | 920 | 896 | **In$_2$Se$_3$** | 40.6 | 59.4 |
| VIII | 700 | 680 | 716 | 726 | **In$_2$Se$_3$** | 42.1 | 57.9 |
| IX | 700 | 650 | 482 | 500 | **In** | 86.4 | 13.6 |